\documentclass[useAMs,a4paper]{mn2e}
\usepackage{savesym}
\usepackage{graphicx}
\expandafter\let\csname equation*\endcsname\relax
  \expandafter\let\csname endequation*\endcsname\relax 
\usepackage{subfig}
\usepackage{amsmath}
\usepackage{amssymb}
\usepackage{verbatim}
\usepackage[yyyymmdd,hhmmss]{datetime}
\usepackage{array}
\usepackage{times}
\usepackage[total={17.8cm,24.0cm},centering]{geometry} 
\usepackage{color}

\newcommand{\be}{\begin{equation}}
\newcommand{\beq}{\begin{equation}}
\newcommand{\ee}{\end{equation}}
\newcommand{\eeq}{\end{equation}}
\newcommand{\eea}{\end{eqnarray}}
\newcommand{\bea}{\begin{eqnarray}}

\newcommand{\dd}{\partial}

\newcommand\W {{W^r_{\ \phi}}}

\title[Relativistic thin discs with finite ISCO stress]{Evolution of relativistic thin discs with a finite ISCO stress: \\ II.  Late time behaviour}
\author  [Andrew Mummery, Steven A. Balbus]{Andrew Mummery\thanks{E-mail: andrew.mummery@physics.ox.ac.uk}, {Steven A. Balbus\thanks{E-mail:
steven.balbus@physics.ox.ac.uk}}
\\
Oxford Astrophysics, Denys Wilkinson Building, Keble Road, Oxford, OX1 3RH, United Kingdom}
\begin{document}

\date{}

\pagerange{\pageref{firstpage}--\pageref{lastpage}} \pubyear{2019}

\maketitle

\label{firstpage}

\begin{abstract} 
We present solutions to the relativistic thin disc evolutionary equation using a modified description of the mean fluid flow within the disc. The model takes into account the effects of sub-circular velocities in the innermost disc regions, and resolves otherwise unsustainable behaviour present in simple finite ISCO stress disc models.  We show that the behaviour of a relativistic thin disc evolving with a finite ISCO stress is comprised of three distinct stages which join the ordinarily distinct finite and vanishing ISCO stress solutions into a fully continuous model parameterisation.  The most important prediction of our model is the existence of an intermediate stage of ``stalled accretion'', controlled by a single dimensionless parameter.  The hallmarks of this evolutionary phase appear to have been seen in GRMHD simulations as well as in the late time X-ray observations of tidal disruption events, but dedicated simulations and extended observations are needed for a deeper understanding.   

\end{abstract}

\begin{keywords}
accretion, accretion discs --- black hole physics --- turbulence
\end{keywords}
\noindent

\section{Introduction}
The late time luminosity from the aftermath of tidal disruption events (TDEs) are generally not well-described by standard models, and are therefore a puzzle.  The recent development of a practical theoretical tool -- a fully relativistic thin-disc evolution equation (Balbus 2017) -- allows a simple but powerful approach to the modelling of thermal emission from evolving thin discs around Kerr black holes (Balbus \& Mummery 2018, Mummery \& Balbus 2019; hereafter BM18 and paper I respectively), and may therefore be of some use in understanding the observed properties of TDEs.   Real TDEs will almost certainly have components in addition to a disc (which itself may or may not be thin), and the disc itself may not be centred on the Kerr equatorial plane, so one should not consider this to be more than a baseline model.   Despite its simplicity, this basic model displays unanticipated features which are nicely compatible with observations, and it is for this reason worthy of further development.   

At late times, TDEs are expected to display light curves $L(t)$ that vary as a power law $n$ in time $t$, $L \sim t^n$.  For example, the original Rees (1988) ``fallback'' model predicts $n = -5/3$, while classical disc models exhibit slightly shallower light curves, with $n = -1.19$ (Cannizzo \textit{et al.} 1990).    Recent observations of the late time X-ray luminosity emergent from four well observed sources (Auchettl, Guillochon \& Ramirez-Ruiz 2017; hereafter AGR) show unexpected behaviour, with measured power law indices clustering around $n \simeq -0.75$.    This is in sharp contrast with the predictions of standard models.  

In BM18, we put forward a new disc model that appears to be in better accord with observations. There are two key differences between this disc model and those of Cannizzo \textit{et al.} (1990).  The first is the use of the relativistic Kerr metric in the disc evolution equation, the second is the imposition of a finite stress boundary condition at the innermost stable circular orbit (ISCO).  It was shown that the inclusion of these two effects (the second is particularly important) significantly modify the late time disc behaviour relative to the Cannizzo \textit{et al.}\ (1990) models.   In both calculations, the evolving  Newtonian disc regions may be understood as arising from a superposition of Laplace modes.    Outer modes that are normally discarded in a fully Newtonian model must be retained in a relativistic treatment. This ensures that the modes describing the outer disc and the inner relativistic disc can smoothly join at some intermediate radius. It is the presence,  and eventual dominance,  of these previously neglected modes that causes the modified disc behaviour in a finite ISCO stress disc.   This has stark observational consequences.   For an electron-scattering-opacity disc model, the luminosity at late times follows $L \sim t^{-11/14} \simeq t^{-0.79} $ (paper I), as opposed to the Cannizzo \textit{et al}. (1990) result of $t^{-19/16}\simeq t^{-1.19}$.    The shallower fall off seems to be a better match to observed TDEs.  

The relativistic evolution equation is derived in the form of a background flow plus small perturbing fluctuations (Balbus 2017). It is usually assumed that the background flow is comprised of circular orbit velocities which are taken from exact Kerr metric solutions (e.g. Page \& Thorne 1974).   In paper I, we argued that care is required when extrapolating to very large times these finite ISCO stress solutions.   This is already evident in the shallow $L(t)$ light curves, whose time integrals formally diverge, and is also seen in the radial velocity field, which becomes {\it positive} over much of the inner disc.    Shallow light curves are not unphysical (they are observed!), nor is positive outflow, but these behaviours are not sustainable indefinitely.     The vanishing angular momentum gradient denominator of the relativistic evolution equation, a consequence of using precise circular Kerr orbits, is sensitive even to small modifications.   These modifications may be required to understand the actual ultimate disc behaviour at very late times.   In this paper we present a general and consistent solution of the relativistic thin disc evolution equation with such a modified angular momentum profile, and show that it leads to sustainable behaviour.  

In this quasi-circular orbit model the evolution of a relativistic thin disc with a finite stress at the ISCO follows three distinct accretion stages.  First, there is an initial inflow of material from the bulk of the disc towards the ISCO, which cannot be immediately accommodated by the black hole.    This  results in a build-up of inner disc material, with the corresponding build-up of the inner disc stress giving rise to an extended period of ``stalled accretion''.   
This period of stalled accretion is quite extended, lasting for many viscous timescales, and is therefore important for interpreting TDE observations.   It is during this stage that the disc follows the unmodified circular orbit solutions discussed in paper I.  The associated shallow luminosity fall-off for the finite ISCO stress solutions are a better match to observations of confirmed TDEs when compared with their zero stress counterparts.   

At much later times, the reduction of the fluid angular momentum in the inner disc regions, a result of sub-circular velocities, means that the inner disc material is depleted, and the surface density lowers.   During this final phase of disc evolution, both the finite and vanishing ISCO stress discs behave similarly, with a large and growing zone of accretion.   These solutions are reasonably well-described by the vanishing stress Cannizzo \textit{et al.} (1990) solutions.   It is, however, the intermediate stage of accretion which is likely to be most important for understanding observations. 

The actual duration of this extended period of stalled accretion is controlled by a single dimensionless parameter, which we denote as $\gamma$.  This is the ratio of the difference between the actual angular momentum at the ISCO and the exact circular orbit value, divided by a quantity known as the `net accreted angular momentum', $j_\text{net} \equiv -\W/U^r$ (Noble, Krolik, \& Hawley 2010).  All of the findings of BM18 and paper I are recovered by taking appropriate limits of this $\gamma$ parameter.  In particular, $\gamma \rightarrow \infty$ corresponds to a simple vanishing stress disc, and $\gamma \rightarrow 0$ to the ideal circular motion finite stress discs.  Determining the value of $\gamma$ in realistic discs is therefore a question of some importance, and will be the focus of future dedicated three-dimensional GR MHD experiments.   In this work, we study numerical solutions of the evolution equation for a broad range of values of $\gamma$, focussing on those values which have emerged from currently published simulations.  

The layout of this paper is as follows. In \S 2 we briefly recap the governing equations, notation and the analysis of paper I  which demonstrated the requirement for a modified angular momentum gradient. In \S 3 we lay-out and solve our refined $\gamma$-disc model, a disc evolving with a modified inner angular momentum profile and a finite ISCO stress, the solutions are found by employing the Laplace mode matching technique of BM18. These refined models make a number of predictions about the properties of evolving thin discs. In \S 4 we present fiducial numerical solutions of our refined disc model, verifying the predictions of the Laplace mode analysis. In \S 5 we highlight avenues for future testing of our model, and recap current evidence for a stalled accretion phase. 

\section{Analysis}\label{analysis}

\subsection{Review of governing equations and notation}
We seek the evolution of the azimuthally-averaged, height-integrated disc surface density $\Sigma (r, t)$, using cylindrical Boyer-Lindquist coordinates for a Kerr disc: $r$ (radial), $\phi$ (azimuthal), and $z$ (vertical).   The contravariant four velocity of the disc fluid is denoted $U^\mu$; the covariant counterpart is $U_\mu$.  The specific angular momentum corresponds therefore to $U_\phi$, a covariant quantity.     There is an anomalous stress tensor present, $\W$, which is assumed to be due to low-level disk turbulence.   This a measure of the correlation between the fluctuations in $U^r$ and $U_\phi$ (Balbus 2017), and in practice will include correlated magnetic fields.  As the notation implies, this is a mixed tensor.   We shall work with the evolution equation in its most compact form (paper I), which describes the evolution of the quantity 
\beq\label{yy}
\zeta \equiv \sqrt{g}\Sigma \W /U^0 =  r \Sigma \W /U^0 ,
\eeq
where $g>0$ is the absolute value of the determinant of the (mid-plane) Kerr metric tensor $g_{\mu\nu}$.  For our chosen coordinates, $\sqrt{g}=r$.   The ISCO radius is denoted as $r_I$, and $x\equiv r-r_I$.  

The governing equation for the evolution of the disc is then given quite generally by (Eardley \& Lightmann 1974; Balbus 2017):
\beq\label{eq}
{\dd\zeta\over \dd t} = \mathcal{W} {\dd\ \over \dd r} \left({U^0 \over U'_\phi}    {\dd\zeta \over \dd r}\right) ,
\eeq
where $\mathcal{W}$ is the stress-like quantity 
\beq
\mathcal{W}(r,\zeta) \equiv  {1 \over (U^0)^2} \left(\W + \Sigma {\dd\W\over \dd \Sigma} \right)  .
\eeq
This could in principle depend on $\zeta$ itself (e.g., in an $\alpha$-disc parameterisation), and the general disc evolution equation is therefore non-linear.  Here and throughout, the primed notation $'$ denotes an ordinary derivative with respect to $r$.   We shall also require the expression for the second-order mean drift velocity of the disc gas.  This is given by the equations of mass and angular momentum conservation (Balbus 2017, paper I) :
\beq\label{radvel}
U^r = - {\W \over U_\phi '} {\zeta ' \over \zeta} .
\eeq

\subsection{Breakdown of mean circular motion approximation in the presence of a finite ISCO stress}
It is important to consider carefully what  is meant by the 4-velocities $U^\mu$ which appear in the evolution equation (\ref{eq}).    This equation is derived perturbatively, with the exact fluid 4-velocity $u^\mu$ split into a mean component $U^\mu$ and a perturbation $\delta U^\mu$. The sole conditions placed upon the respective 4-velocity components are the following scaling relationships
\beq\label{assumption}
\delta U_\phi \ll U_\phi, ~~~ U^r \ll \delta U^r \sim \delta U_\phi / r \ll r U^\phi .
\eeq
The underlying assumption expressed by these relationships is that the fluid  external to the ISCO moves, at leading order, on circular geodesic orbits.   At first order fluctuations appear, and finally at second order there is a radial drift velocity, produced by the turbulent stress of correlated first order fluctuations. 
This formalism implies, in particular, that the zeroth order mean fluid motion is independent of the properties of the disc stress.   For such circular orbits, the angular momentum gradient $U_\phi'$ is zero at the ISCO.   Expanded and rearranged, equation (\ref{eq}) may be written 
\beq
U_\phi ' \left[ U_\phi ' \dot \zeta - \mathcal{W} U^0 \zeta '' + \mathcal{W} \left(U^0\right) ' \zeta' \right] + \mathcal{W} U^0  U_\phi '' \zeta ' = 0.  
\eeq
Therefore, in a finite ISCO-stress disc, the radial derivative of $\zeta$ must vanish along with the angular momentum gradient at the ISCO. 

In paper I we found that the resulting disc solutions were unsustainable at large times.  In these models, $\zeta (r_I)$ eventually became a global maximum, leading (via eq.\ [\ref{radvel}]) to radial outflow external to $r_I$ (but also the observed shallow light curves).   In a real disc, this would have to self-regulate.  Eventually there would be a fall in the disc stress at the inner edge, when the material interior to $r_I$ runs out.    But a decrease in the inner disc stress would also lower $\zeta$, leading to a recommencement of accretion.   In the presence of a finite ISCO stress, therefore, the usual assumption (e.g. Page \& Thorne 1974) that the properties of the mean fluid flow are those of precisely circular motion, independent of the turbulent disc stress, needs to be revisited.   

We are motivated to modify the angular momentum gradient of the fluid elements allowing for the fact that it is not precisely zero at the (formal) ISCO, owing to radial flow.    This modification is physically reasonable.   Finite inward radial velocities generally will reduce their angular momentum below that required of a circular orbit.  Ordinarily, this small change in the angular momentum is a second order effect and therefore negligible.    But at the ISCO a small change in the vanishing angular momentum gradient denominator represents an important deviation.    We shall accordingly investigate the simplest possible self-consistent thin disc model incorporating this:  the evolution of an initially localised disc evolving with a finite ISCO stress and a non-zero angular momentum gradient at the ISCO.

\section{a modified disc model}\label{newmodel}
\subsection{ Angular momentum profile}

In the fundamental disc equation, only two mean flow quantities appear, $U^0$ and $U_\phi '$.  The gross properties of the disc are quite insensitive to small deviations in $U^0$.   The key modification required is then in the gradient $U'_\phi$.    A generic change to the angular momentum gradient takes the form
\beq\label{newcirc}
U_\phi ' = \left(V_\phi^c\right) ' + \epsilon (r) ,
\eeq
where we use the notation $U_\phi$ to indicate the mean fluid angular momentum entering equation (\ref{eq}), and $V_\phi^c$ is the corresponding angular momentum of an exact circular orbit. The function $\epsilon(r)$ is in our model somewhat arbitrary.   However, it is constrained by two simple properties: (i) it reaches a maximum at the ISCO, and (ii) it vanishes for $r \gg r_I$.  (The idea is that ideal circular orbits are a good approximation to the mean fluid dynamics at large radii, and become progressively worse as the ISCO is approached.)      It is convenient to adopt an exponential form 
\beq\label{epsdef}
\epsilon(r) = \frac{\Delta j}{L} \exp\left(-x/L\right) , ~~x \equiv r - r_I.
\eeq
We consider $x>0$ only.   For $x < 0$, the radial velocity of the disc material increases rapidly inwards and the disc density will see a correspondingly rapid decrease.  
The  parameter $L$ is the length scale over which the deviations from a circular orbit is important.   In this paper, we adopt $L = 10 r_g$,  so that the change from circular orbits to in-spiralling orbits in the stable disc regime is a gradual one.  The second parameter ($\Delta j$) is a measure of the magnitude of the  deviation of the angular momentum of the material at the ISCO from that of the circular orbit angular momentum.  This is seen by integration of equation (\ref{newcirc})
\beq\label{angprofile}
U_\phi = V^c_\phi - \Delta j \exp\left(-x/L\right).
\eeq
At the ISCO $r_I$ itself,
\beq
U_\phi(r_I) = V^c_\phi(r_I)  - \Delta j .
\eeq
The quantities $\epsilon(r)$ and $\Delta j$ are merely model parameters here.  We adopt a simple and convenient functional form so as to understand the sensitivity of the disc solutions to a modified angular momentum gradient.   A more precise form of the disc angular momentum profile is best determined by direct numerical simulation.  Our approach is thereby testable.  

\subsection {ISCO boundary condition}
In BM18, we argued that the inner matching constraint is that the outer Newtonian $\zeta$ modes must match smoothly onto the local ISCO Ai$'$  function, selected for the property that it decays exponentially interior to the ISCO radius.   In this work, with a finite $U'_\phi(r_I)$, we adopt a more utilitarian approach, closer in spirit to specifying the accretion rate $\dot M$ in an equilibrium disc model.  In fact, following equation (\ref{radvel}), there is only one possible boundary condition which conserves angular momentum within the disc:
\beq
{\zeta' \over \zeta} =  -  {U_\phi ' U^r\over W^r_{\, \phi}} \qquad ({\rm at\ }r=r_I).
\eeq
The (density weighted) ratio $-(\rho W^r_{\, \phi})/(\rho U^r)$ is denoted $j_{\text{net}}$ by Noble \textit{et al}.\  (2010), which in our one-dimensional model we identify with $-\W/U^r$.    It may be readily extracted from numerical simulations.   With our chosen functional form of the angular momentum gradient (equation \ref{epsdef}), the boundary condition becomes 
\beq\label{gammaBC}
{\zeta' \over \zeta} = -  {\Delta j\over L}{ U^r\over W^r_{\, \phi}} =
 {{1}\over{L}}  {\Delta j\over j_{\rm net}  } \equiv {\gamma\over L} .
\eeq
In specifying the dimensionless parameter $\gamma$ (the ratio of the angular momentum circular orbit deficit to $j_{\rm net}$), we impose our ISCO boundary condition for $\zeta$.     The exact circular orbit limit corresponds to $\gamma=0$.   We will see that the size of $\gamma$ profoundly influences the disc evolution in these models. 

{For the remainder of this paper, we treat the parameter $\gamma$ as a constant and examine its effect on an evolving thin disc.  This is clearly a simplifying assumption:  it may be that $\gamma$ is sensitive to the disc stalling and changes its value accordingly.   This may arise, for example, if there were a drop in the magnitude of the ISCO stress caused by the drainage of material interior to the ISCO, which is then not replenished by exterior material due to accretion stalling.    Given that none of the parameters $\W$, $\Delta j$ and $U^r$ can be determined from first principles, the actual behaviour of $\gamma$ is best addressed by numerical simulations.   Even restricted by this simplification, these constant $\gamma$ models demonstrate a mathematically rich and astrophysically interesting pattern of behaviour.}

\subsection{Linear Laplace mode analysis} 
The dependence of the disc properties on $\gamma$ can be understood with a normal mode Laplace decomposition, as in BM18.
This assumes a linear governing equation and involves a piecewise smooth analysis of the modal solutions of the disc equations,  joining the near-ISCO strong-field modes to the outer Newtonian modes at an intermediate matching radius.   In the outer regions of the disc, the Laplace mode solution (time dependence $e^{-st}$) will have the general form 
\beq
\zeta(r,s) = c_1(s)\, \zeta_1(r,s) + c_2(s)\, \zeta_2(r,s),
\eeq
where $\zeta_1$ and $\zeta_2$ are linearly independent spatial amplitudes. Crucially, in the relativistic problem, both modes will in general be required to smoothly join onto the near-ISCO solutions.  The amplitude $\zeta_2$ vanishes at the origin along with its gradient, while only the gradient of $\zeta_1$ vanishes at $r=0$.  We shall always write the outer Laplace mode solution in this form: the  `$c_1$' solution is understood to have a vanishing first derivative at the origin, while the `$c_2$' solution vanishes itself at the origin.   The key question, both mathematically and physically, is which of the outer modes dominates within the disc at late times.   It is precisely this question which can be answered using the smooth joining continuity conditions, as they allow the ratio $c_1/c_2$ to be determined.
The calculations of BM18 may also be followed here using our modified angular-momentum-gradient formalism, which of course changes only the form of the inner strong field solutions.  As before, the inner solution matching strongly modifies the {\em global} behaviour of the disc as $t\rightarrow\infty$. 

To make progress we treat the simplest analytic case in which the turbulent stress is a constant ($w$) at all radii.   We use the evolution equation in reduced form, which retains the physical content of the full equation, but in a mathematically simpler form:
\beq
{\dd y\over \dd t}= w{\dd\ \over \dd r} \left( {1\over U'_\phi}{\dd y\over \dd r}\right) .
\eeq
{This reduced form amounts to setting $U^0 = 1$ in equation (\ref{eq}).  Since $U^0$ is smooth, non-vanishing, and varies only modestly over the domain of interest,  all essentials are retained by this convenient simplification.   (This has been verified numerically.)   To distinguish the reduced form from the full equation we denote the dependent variable $y$ as a proxy for $\zeta$.  }
With 
\beq 
U_\phi ' = {{\Delta j}\over{L}}\exp(-x/L),
\eeq
the local form of this equation near the ISCO is
\beq
\frac{\partial y}{\partial t} = \frac{w}{L \Delta j } \frac{\partial}{\partial X}\left( e^X \frac{\partial y }{\partial X}\right) ,
\eeq
where we have introduced $X \equiv (r-r_I)/L$, a dimensionless length. 
For stable Laplace modes ($y \sim e^{-st},~ s > 0$), this becomes
\beq
\frac{\partial}{\partial X}\left( e^X \frac{\partial y }{\partial X}\right) = - s \frac{L \Delta j }{w} y \equiv - b^2 y .
\eeq
This equation has exact solutions of the form 
\beq\label{exinn}
y =  e^{-X/2} \left[ c_3 J_1\left( 2b e^{-X/2} \right) + c_4 Y_{1}\left( 2b e^{-X/2}\right) \right] ,
\eeq
where $J_1$ is the standard Bessel function of the first kind, and $Y_1$ the corresponding Bessel function of the second kind.  Using standard Bessel function identities, the derivative of $y$ with respect to radius is
\beq\label{derinn}
\frac{\text{d}y}{\text{d}r } = - \frac{b e^{-X}}{L} \left[c_3 J_0\left(2be^{-X/2}\right) + c_4 Y_0 \left(2be^{-X/2}\right) \right] .
\eeq
The outer Newtonian equation is completely unchanged (BM18).   In this region, 
\beq
U_\phi ' \approx {{1}\over{2}} \sqrt{{{GM}\over{r}}},
\eeq
so that
\beq\label{toy4}
{\dd y \over \dd t} =  {2w\over \sqrt{GM}}{\dd\ \over \dd r}\left( r^{1/2} {\dd y \over \dd r}\right).
\eeq
The (Newtonian) modal solutions of this equation, denoted $y_N$, are given by (BM18)
\beq
y_N =  r^{1/4}\left[ c_1 J_{-1/3}\left(pr^{3/4}\right) +c_2 J_{1/3}\left(pr^{3/4}\right)\right] ,
\eeq
where $J_{\pm1/3}$ are the usual Bessel functions of fractional order,  and
\beq
p^2 \equiv {\frac{s\sqrt{GM}}{3w}}.
\eeq 
The gradient of $y_N$ is 
\beq\label{grdi}
\frac{\text{d}y_N}{\text{d}r} = \frac{3p}{4} \left[c_2J_{-2/3}\left(pr^{3/4}\right) - c_1 J_{2/3}\left(pr^{3/4}\right) \right] .
\eeq

The smooth matching condition is that at some intermediate radius $r_m$ the inner and outer modes have the same amplitude $y(r_m) = y_N(r_m)$ and the same gradient $y'(r_m) = y_N'(r_m)$.  The key quantities to determine are the coefficients $c_1$ and $c_2$. The matching conditions, presented with the respective Bessel function arguments suppressed (note that both $p$ and $b$ scale as $s^{1/2}$), are: 
\begin{align}
c_1 J_{-1/3} + c_2 J_{1/3}  &= A(s) \label{jc1} , \\
 c_1 J_{2/3} - c_2 J_{-2/3}  &= B(s) \label{jc2} . 
\end{align}
Here we have defined
\begin{align}\label{ABs}
A(s) &=  \frac{e^{-X_m/2}}{r_m^{1/4}} \left( c_3 J_1 + c_4 Y_{1} \right), \\
B(s) &= \frac{4be^{-X_m}}{3pL} \left(c_3 J_0 + c_4 Y_0  \right) .
\end{align}
The solutions of equations (\ref{jc1}) and (\ref{jc2}) are
\begin{align}
c_1 &= [A(s) J_{-2/3} + B(s) J_{1/3}]/ {\text{Wr}}, \\
c_2 &= [A(s) J_{2/3} - B(s) J_{-1/3}]/ {\text{Wr}} ,
\end{align}
where ${\rm Wr}$ is a Wronskian formed from the fractional Bessel functions:
\beq
\text{Wr} \equiv J_{1/3} J_{2/3} + J_{-1/3} J_{-2/3} = {{\sqrt{3}} \over {\pi pr_m^{3/4}}} \propto s^{-1/2} .
\eeq
We next impose the  ISCO boundary condition 
\beq
\frac{y'}{y} = \frac{\gamma}{L}.
\eeq
Substituting from equations (\ref{exinn}) and (\ref{derinn}) and rearranging, we find 
\beq\label{c3c4}
\frac{c_3}{c_4} = - \left( \frac{b Y_0 + \gamma Y_1}{b J_0 + \gamma J_1} \right) .
\eeq
At late times, the $s \rightarrow 0$ behaviour of our solution is important.  The  Bessel functions have the following expansions for small $z$:   
\begin{align}
Y_1(z) &\sim 1/z, \\
Y_0(z) &\sim \ln (z), \\
J_1(z) &\sim z, \\
J_0(z) &\sim \text{constant} ,
\end{align}
and more generally
\beq
J_{ \nu}(z) \sim z^{ \nu} .
\eeq
Equation (\ref{c3c4}) shows that 
$c_3/c_4\sim s^{-1}$ in the  $s \rightarrow 0$ limit.   This then implies 
$A(s)/B(s)\sim s^{1/2}$, 
and thus, in the $s \rightarrow 0$ limit,
\beq
c_2/c_1 \sim s^{-1/3}.
\eeq
The coefficient $c_2$ therefore dominates at extremely late times.   This is the late time behaviour that arises in standard vanishing ISCO stress models, but now we see that it holds even in the presence of stress!   {\it This is a very interesting result:}  the reduction of the average disc angular momentum to a value below that of a circular orbit has lead to a qualitative change in the {\em outer} disc behaviour as $t\rightarrow \infty$. 

How do we reconcile this finding with the conclusion of BM18, that $c_1$ behaviour, with its shallow light curve fall-off, follows from the presence of an ISCO stress?   The answer comes down to a delicate asymptotically ordering of time scales, and the smooth matching conditions are once again key.   The point is that equation (\ref{c3c4}) introduces another asymptotic scale into the set of equations:  the relative size of $b$ and $\gamma$. The coefficient $b$ is an $s$-dependent quantity, and so the ratio $\gamma/b$ will be different for each mode. 

Consider the case $\gamma/b \rightarrow 0$, the opposite of a late time, small $s$, constant $\gamma$ limit.    The case we now consider will be appropriate for times restricted to be earlier than a characteristic time scale which is derived below, or a vanishingly small $\gamma$.   In this limit, $c_3/c_4 \rightarrow - Y_0/J_0$, implying that $B$ vanishes while $A$ remains finite.    Finite $s$, vanishing $B$ modes satisfy 
\beq
c_1 \propto  {J_{-2/3}\left[p(s)r_m^{3/4}\right]}, ~ c_2 \propto {J_{2/3}\left[p(s)r_m^{3/4}\right]} .
\eeq
The large argument limit of a Bessel function is 
\beq
J_\nu(z) = \sqrt{\frac{2}{\pi z}} \cos \left[ z - \left(\frac{2\nu + 1}{4}\right)\pi\right] + O(1/z^{3/2}) \, ,
\eeq
meaning that in the $p(s) r_m^{3/4} \gg 1$ limit, $c_1/c_2$ is of order unity:
\beq
\frac{c_1}{c_2} \simeq \frac{\cos\left[p(s)r_m^{3/4} - 7\pi/12\right] }{ \cos\left[p(s)r_m^{3/4} + \pi/12\right] }\sim O(1),
\eeq 
whereas when $s$ is sufficiently small that $p(s)r_m^{3/4} \ll 1$,
\beq
c_1/c_2 \sim s^{-2/3} \gg 1.
\eeq
Therefore,  modes with $b \gg \gamma$ will have $c_1$ solutions comparable to, or dominating, the $c_2$ modes.   This inequality 
gives an effective time scale over which these modes will be important (all modes are of course ultimately suppressed by a factor $e^{-st}$).   These modes will be important up until a time $t_c \equiv 1/s_c$, where $s_c$ is defined through the modal condition $\gamma / b(s_c) \simeq 1$:
\beq
\gamma\simeq b(s_c) = \sqrt{ \frac{s_c L \Delta j }{w} } =\sqrt{ s_c L\gamma\over |U^r|},
\eeq
where we have used equation (\ref{gammaBC}).   Solving the above for $t_c=1/s_c$ in terms of the viscous time scale appropriate for the inner disc regions, $t_v \equiv L/|U^r|$, we obtain an expression for the critical time scale: 
\beq\label{t_scale}
 t_c \simeq \frac{t_v}{\gamma} .
\eeq
This means that for times $t \lesssim t_c$ the (shallow fall off) $c_1$ modes will be of at least comparable importance to the (steeper) $c_2$ modes; as we pass through $t=t_c$ the latter dominates.   The key point is that if $t_c$ is sufficiently large,  there will be a extended period over which the $c_1$ solution dominates within the disc, and the accretion is stalled.    

The results of BM18 --- that for circular orbits the (steep-falling) $c_2$ solutions dominate when the ISCO stress is zero, and that the (shallow-falling) $c_1$ solutions dominate when the ISCO stress is finite --- can be fully recovered by taking appropriate limits of the time parameter $t_c$.   For vanishing ISCO stress $\gamma \rightarrow \infty, ~t_c \rightarrow 0$ and the period over which $c_1$ dominates is vanishingly brief. While for \textit{any} non-zero $\gamma$ the $c_2$ solution will eventually dominate,  a finite ISCO stress with perfect circular orbits corresponds to the $\gamma \rightarrow 0$ limit, in this limit $t_c \rightarrow \infty$ and the $c_1$ solutions will dominate at late times.

From an observational standpoint,  the most important prediction of the existence of the stalled accretion phase is that the luminosity from transient astronomical disc sources should  follow much more shallow decay profiles than they would from classical vanishing stress disc modelling.   This was the major observational prediction (or ``post-diction'') of BM18.   The quantity of interest is the power law index, $n$---now perhaps best denoted as $n(t)$--- a quantity that would be measured at each time in the light curve decay from a fit of the form $L \sim t^{n}$.   Classical vanishing and finite ISCO stress models based on circular orbits make clean predictions of the power law index $n$ at late times in the evolution of the disc.   Here the behaviour is richer.  During the stalled phase of accretion the power law index will be close to that of the $c_1$ solutions,  $n \sim -0.7$, while at very late times the power law will be given by the $c_2$ solutions, typically $n \sim -1.2$ (BM18, paper I, Cannizzo \textit{et al}. 1990). The power law index evolves as time progresses, from smaller values less than unity in the stalled phase, to larger values greater than unity in the very-late time approach to the steady state. 

To summarise:  time-dependent accretion  is generally a {\em three stage} process, controlled by what we have identified as the $\gamma$ parameter.   The early behaviour of a spreading accretion disc is insensitive to the stress and angular momentum modelling.   Most of the disc material spreads rapidly inwards toward the ISCO, while the outermost regions move outward to take up the angular momentum of the inward bound material.  This inward-moving material forces a density (and dynamical stress) build up in the inner disc regions, leading to a period of dominance of the $c_1$ solutions,  characterised by stalled accretion and shallow light curves.    These may be quite extended in time.   Eventually, the ongoing slow depletion of the inner disc material lowers the disc density and dynamical (as opposed to kinematical) stress in this region, and a growing dominance of the (steep light curve) $c_2$ solutions at the latest times. 

\section{Numerical $\gamma$-disc solutions }

As a verification of the \S 3 analysis, we have calculated numerical solutions of the evolution equation (\ref{eq}), with angular momentum profiles given by equation (\ref{newcirc}).  The mathematical problem to be solved is the evolution of a very compact ring, in effect a Green's function solution.  In this case, even the evolution from a numerically discrete, single-grid-point delta function is stable, and rapidly smooths.    Consistent with the above analysis, we find that accretion in a finite ISCO stress thin disc follows the three stage pattern outlined above.
 
\subsection{Fiducial $\gamma$-disc model: $\W = w,$ $a = 0,$ $L = 10r_g$}
The $\gamma$ parameter may be estimated from numerical simulations.   Noble \textit{et al}.\ (2010) ran a series of disc simulations for a range of disc thicknesses and initial magnetic field geometries.    Each of these simulations found {\it i)} a finite stress at the ISCO; {\it ii)} the near-ISCO local angular momentum less that of a circular orbit, and {\it  iii) } a $j_{\rm net}$ ISCO angular momentum flux also below that of a  circular orbit.  Typical values, normalised in units where $GM = c = 1$,  were $\Delta j \simeq 0.1 - 0.3$,  and  $ j_{\rm net}\simeq 3-3.3$ (see figs. 14 and 15 of Noble {\it et al.} [2010]).   Given the ample uncertainties, we have therefore examined a broad range of (dimensionless) $\gamma \sim 0.01 - 0.1$.   

In our fiducial model, the the simplest possible finite ISCO turbulent stress model is assumed: $\W = w = $ constant everywhere.   To implement a vanishing stress boundary condition at $r=r_I$ (which is only necessary for the $\gamma \rightarrow \infty$ limit), for $r\le 10 r_g$ the stress departs from its otherwise constant $w$ value following the prescription
\beq
\W = w { ( r-r_I)^2 \over (10r_g-r_I)^2 }.
\eeq
We assume a Schwarzschild black hole ($a = 0$), and the initial condition is a single grid point delta function ring, initially located at $r_0 = 15 r_g$.  This is a nominal tidal radius taken from Rees (1988)\footnote{This was mistakenly identified in BM18 as the tidal radius of a solar mass star around a $10^6 M_{\sun}$ black hole:  it is in fact the tidal radius of a solar mass star about a $4 \times 10^6 M_{\sun}$ black hole. }, $\sim 2.5$ times the ISCO radius of a Schwarzschild hole.  The natural dimensionless time for the viscous spreading of the disc is (see BM18, Appendix 1):  
\beq
\tau \equiv \frac{9wt}{2\sqrt{GMr_0^3}}.
\eeq
(Note that this differs from the $\tau$ used in paper I, \S 2.5, which was normalised for other purposes.)
Finally,  we use equation (\ref{gammaBC}) to set the inner boundary condition.   

In order to follow the rich behaviour of both the light curves and the power law indices themselves in this modified disc model, it is important to be able to convert the formal viscous timescale of these solutions to one that it is observationally convenient.   We adopt  the viscous timescale from $\alpha$ disc theory, noting that $\Omega$ has the same form in both Newtonian gravity and Schwarzschild geometry:
\beq
t_v = \alpha^{-1} (H/R)^{-2} \, \Omega^{-1}.
\eeq
Using the following typical values appropriate for comparison to a generic tidal disruption event light-curves: $H/R = 0.02$, $\alpha = 0.1$, $M_{BH} = 4 \times 10^6 M_{\sun}$ and $ r_0 = 15 r_g $ (as above), leads to $t_v = 330$ days,  which fixes our $w$ parameter.  The time axis of these plots is accordingly plotted from the numerical $\tau$ using the relationship $t = 330 \tau$ days.  It should be regarded as representative only.   

The disc is truncated at its inner edge by the ISCO, at which point the gas flows rapidly inward,  and the evolution equation (\ref{eq}) no longer accurately models the disc dynamics.  Since the disc surface density will then fall off sharply, we shall assume that any interior emission will not significantly alter the integrated disc luminosity.

\begin{figure}
\includegraphics[width=.5\textwidth]{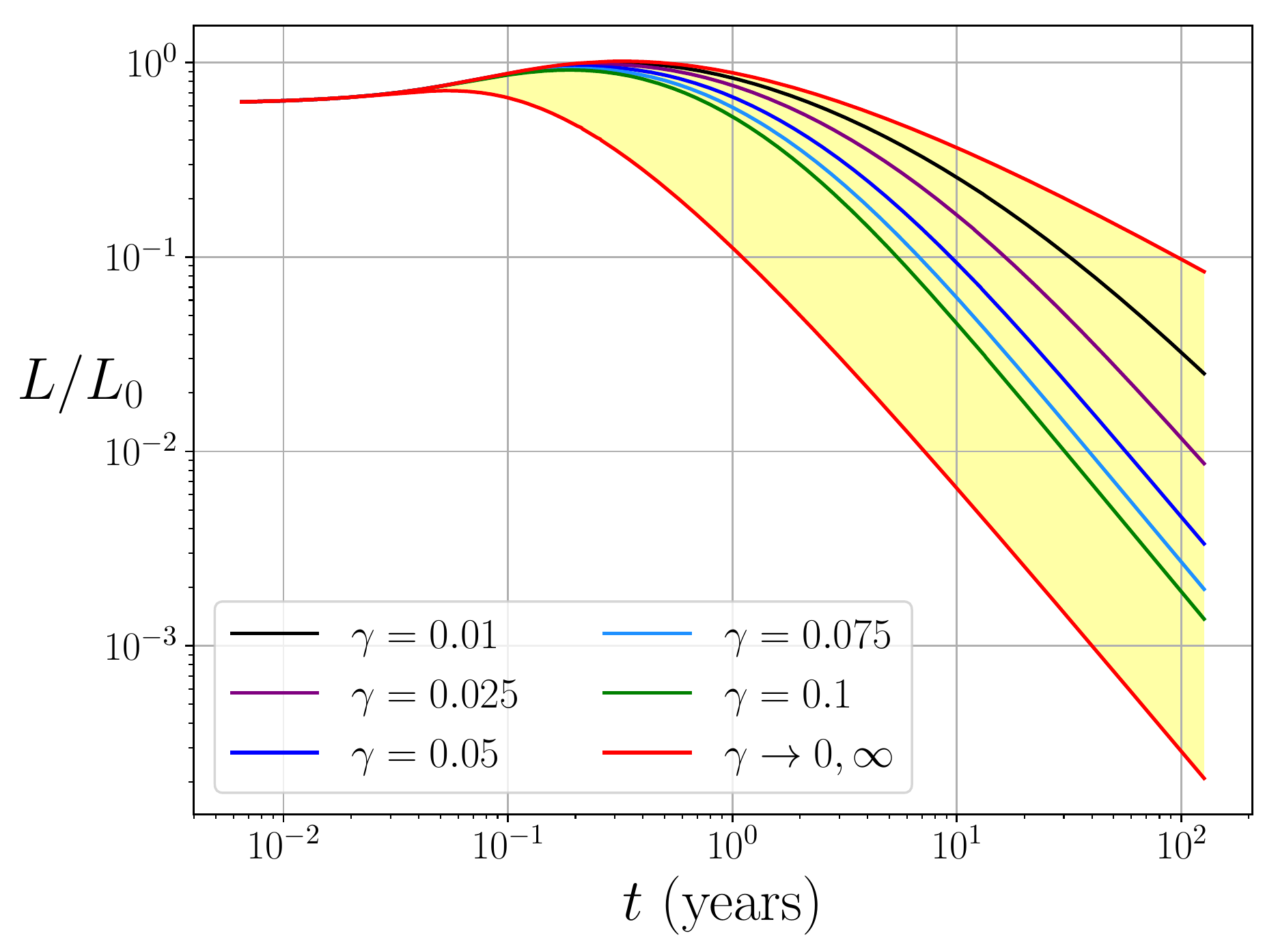} 
\caption{The  evolving luminosity for a series of $\gamma$-discs in a Schwarzschild spacetime. The turbulent stress is given by $\W = w$, a constant. The light curves are normalised by the peak luminosity of the brightest disc studied.  As predicted, the light-curves of these modified $\gamma$-disc models are bounded by the simple vanishing stress (lower red) and finite stress (upper red) light curves.  }
 \label{L_compare}
\end{figure}

\begin{figure}
\includegraphics[width=.5\textwidth]{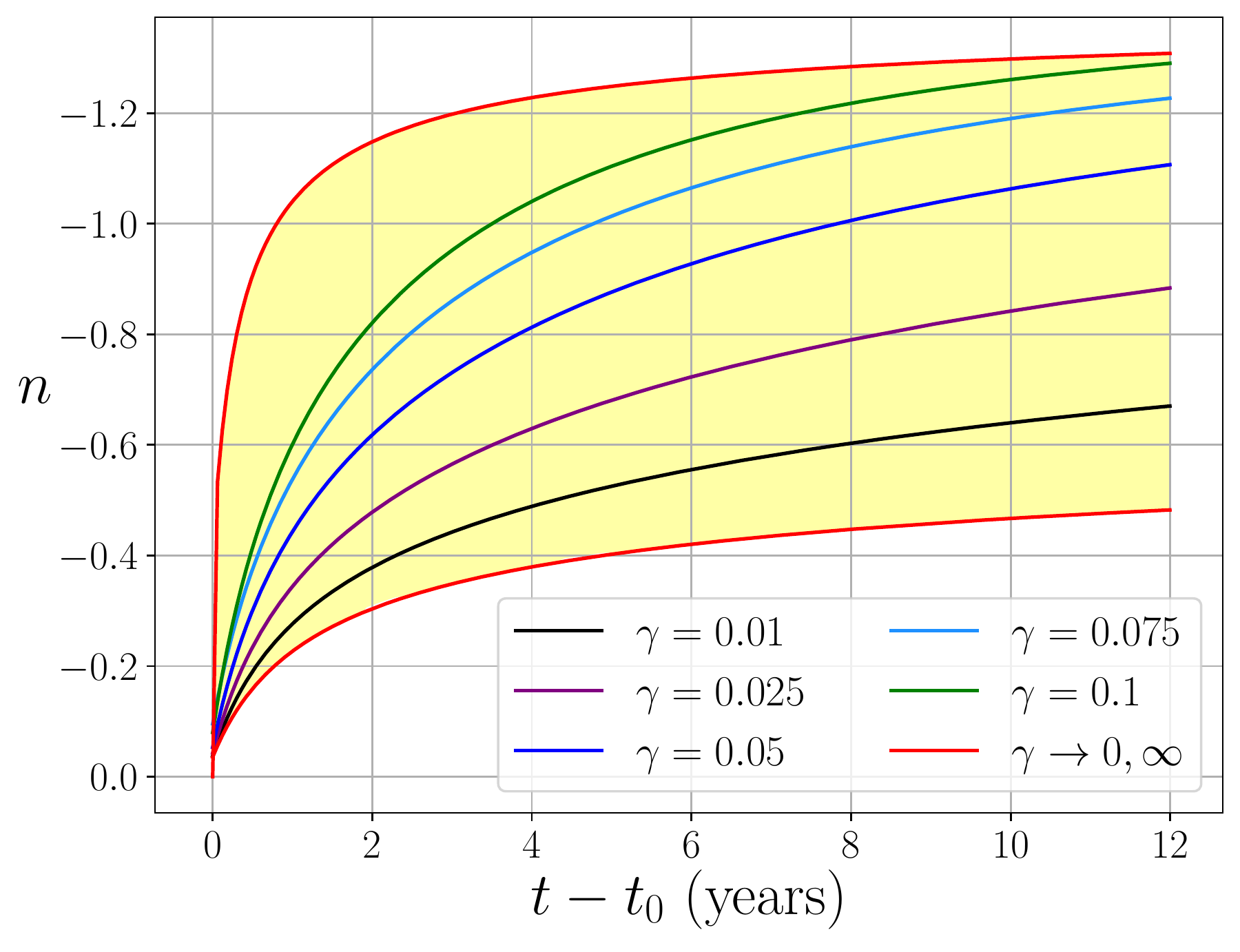} 
\caption{The evolving luminosity decay index $n$, a best fit to $L \sim t^{n}$ for each light curve in Fig.\  (\ref{L_compare}), plotted as a function of time since the peak of the respective light-curves, $t_0$. The upper red curve now corresponds to a vanishing ISCO stress, whereas the lower red curve corresponds to the paper I finite stress model. }
 \label{n_compare}
\end{figure}

\subsubsection{Luminosity evolution }
Fig.\  (1) shows the evolving luminosity for a sequence of discs, differing only by their respective $\gamma$ values, which were motivated by those inferred from GRMHD simulations.   For reference, we also show the light-curves of the two simpler { BM18} disc models, which together bound the finite $\gamma$ runs from above and below.  The lowest red curve corresponds to a vanishing ISCO stress disc ($\gamma \rightarrow \infty$), while the uppermost red curve corresponds to ideal circular motion in a finite ISCO stress disc  ($\gamma = 0$).  
The light curves for a series of discs differing only by their $\gamma$ value should lie in the yellow shaded region.

Fig.\  (\ref{n_compare}) shows the effective local luminosity decay index $n= d\ln L/d\ln t$ as a function of time, starting with the light-curve peak value  and following for an extended period of time.   Note that the $\gamma \rightarrow \infty$ vanishing stress case evolves very rapidly over a matter of months, and quickly achieves $|n| > 1$.   By contrast, the light curve emergent from any finite-$\gamma$ disc has a much more extended period, characterised by a smaller luminosity decay index, of order $n \sim {-0.7}$ in the first several years.  This is the hallmark of an extended period of stalled accretion, and when filtered by bandpass, an observational prediction.     All of the non-zero-$\gamma$ decay-index curves eventually approach that of the vanishing stress solution ($|n|>1$) at the very latest times, which lie off scale to the right of the $t-t_0$ values plotted.

\subsubsection{Radial fluid flow evolution}
The nature of the radial flow in the inner disc determines which of the modal solutions dominates in the outer disc, and therefore much of the gross dynamical behaviour of the disc itself.   To track the behaviour of the radial flow, the radius of zero radial velocity, $R_T$, is calculated at each time step as part of our solution.   Inward of $R_T$ the flow is accreting; outward of $R_T$ it is expanding.  The time dependence of $R_T$ determines the relative importance of the outer modal solutions, since these two solutions are linked to very different late time behaviours for $R_T$.     For the stalled $c_1$ solutions, this is $R_T \rightarrow 0$;  whereas  for the $c_2$ solutions, $R_T \sim t^{2/3} \rightarrow \infty$ (for our constant turbulent stress models).

\begin{figure}
\includegraphics[width=.5\textwidth]{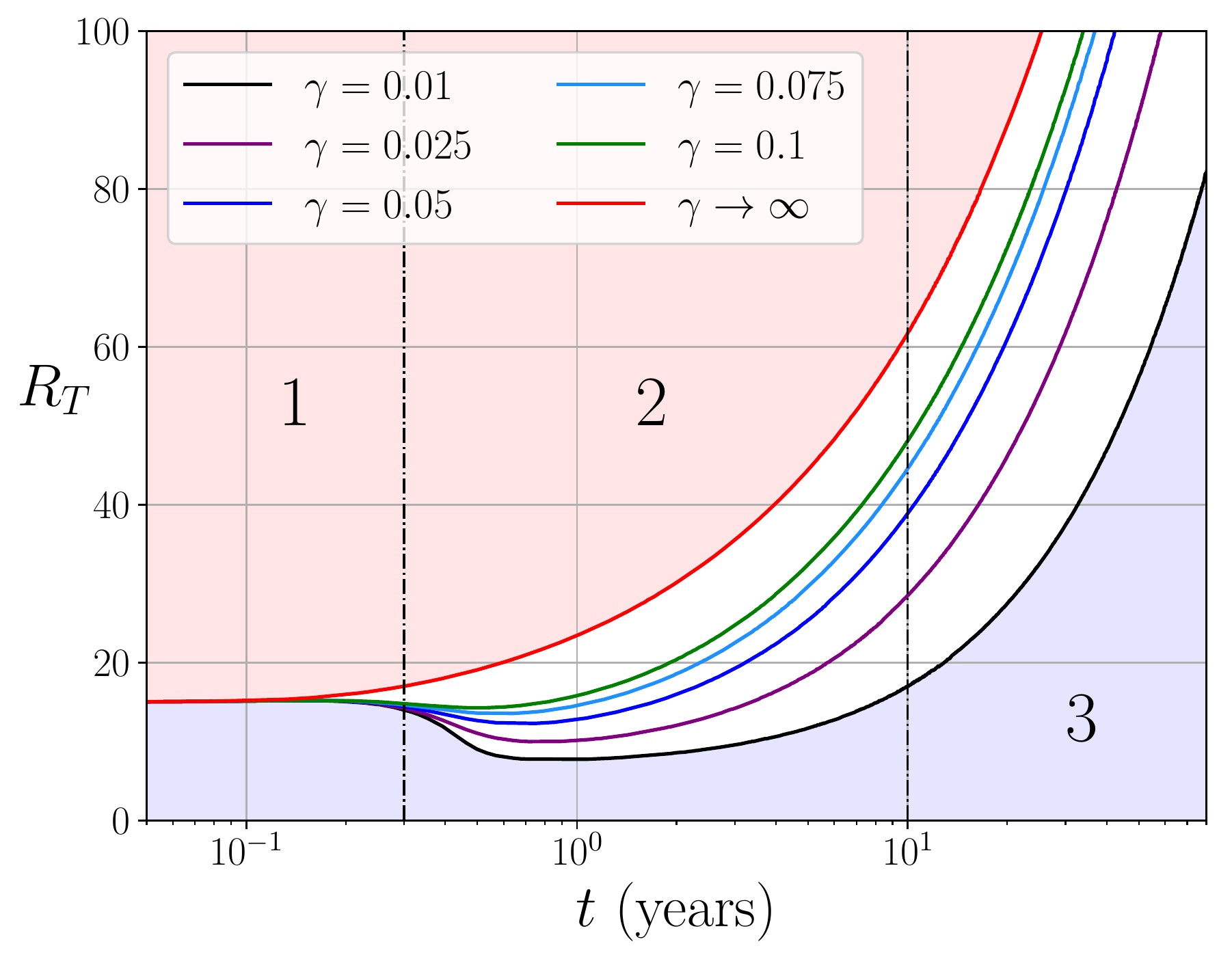} 
\caption{The long-term time evolution of $R_T$, the radius of zero radial velocity, for different values of $\gamma$.  The location $R_T$ separates inflow from outflow in the disc.    Radial outflow is present for $ r > R_T$ and inflow for $r < R_T$.   All discs studied are outflowing in the red shaded region and inflowing in the blue shaded region. 
The three stages of accretion are thus roughly separated by the vertical dashed lines: (1) rapid inflow; (2) stalled accretion; (3) approach to steady state.}
 \label{R_compare}
\end{figure}

\begin{figure}
\includegraphics[width=.5\textwidth]{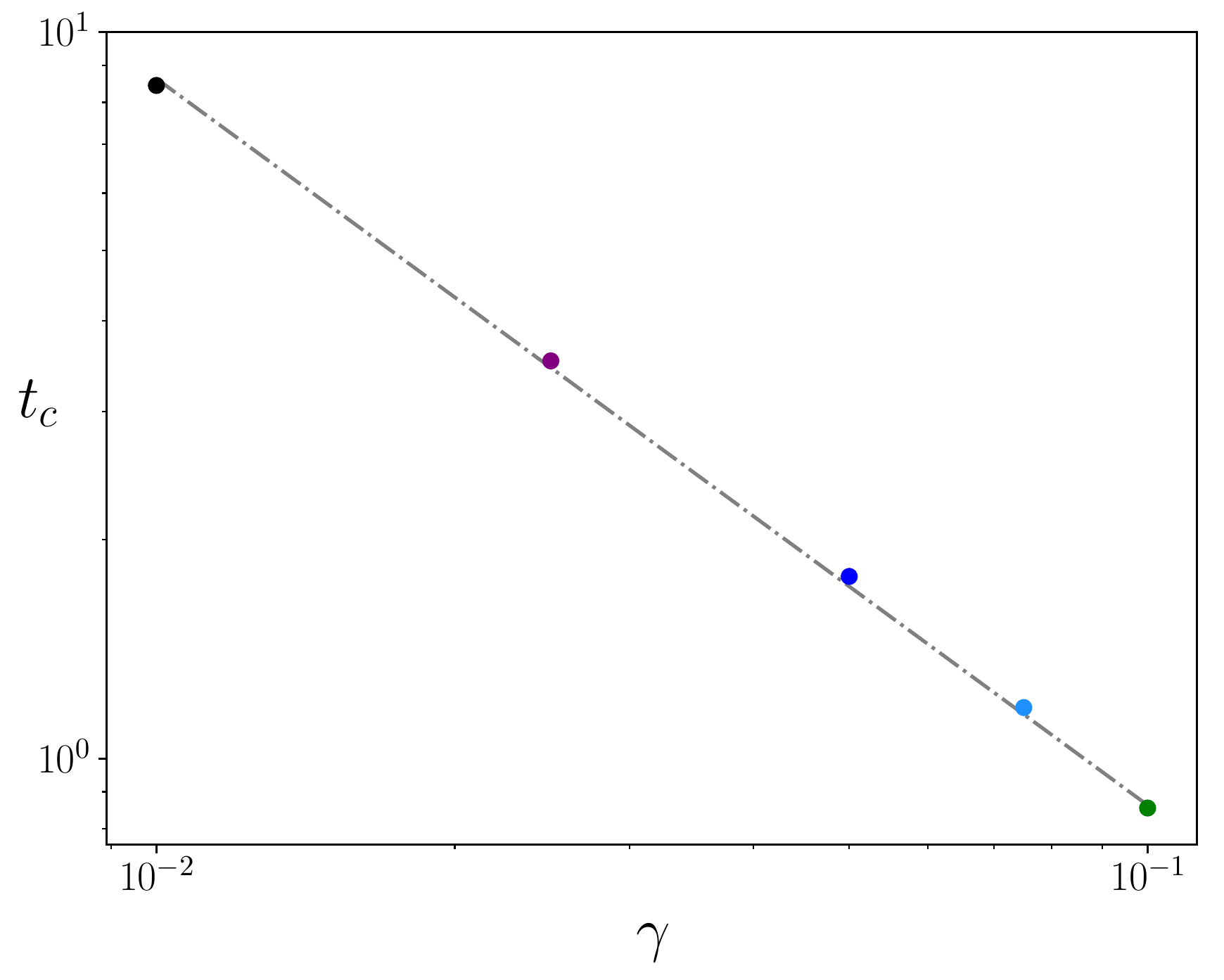}
\caption{The time (in years) required for $R_T$ to begin its upwards trajectory, defined as the time when $R_T$ first becomes greater than $r_0$, plotted as a function of $\gamma$. This quantity is a proxy for the length of the stalled phase of accretion. The grey dashed line is a fit of $t_c \propto 1/\gamma$. This fit is in excellent agreement with the Laplace mode analysis result $t_c \sim 1/\gamma$ equation (\ref{t_scale}).}
\label{tc}
\end{figure} 

The evolution of $R_T$ is plotted in fig.\  (\ref{R_compare}) for five different values of $\gamma$.    Note the logarithmic timescale.  
The vertical black lines separate the figure into the three different disc accretion regimes: rapid inflow (1);  stalled accretion (2); and the approach to the steady state (3).  Inspection of fig.\  (\ref{R_compare}) demonstrates the stalling associated with a finite ISCO stress: while the eventual outward trajectory of $R_T$ is similar for all values of $\gamma$, there is generally a significant time delay before this begins, particularly for the smaller values of $\gamma$.

A measure of the stalling time interval, $t_c$, is defined by the time at which $R_T$ first becomes greater than $r_0$. The value of $t_c$ is plotted in fig.\  (\ref{tc}) for the different $R_T$ evolution profiles in fig.\  (\ref{R_compare}).  We also plot a the best fit curve under the assumption that $t_c \propto 1/\gamma$.    From the Laplace mode analysis of \S\ref{newmodel}, this scaling is how the time over which the $c_1$ solutions dominate the outer disc (the cause of the stalling phase) should indeed vary with $\gamma$.     The zero-ISCO-stress disc has a vanishingly brief period of stalled accretion, as the disc passes directly from evolutionary stages 1 to 3 with $R_T$ continuously increasing.  

\subsubsection{The evolution of $\zeta$}

\begin{figure*}
\centering
\includegraphics[width=\textwidth]{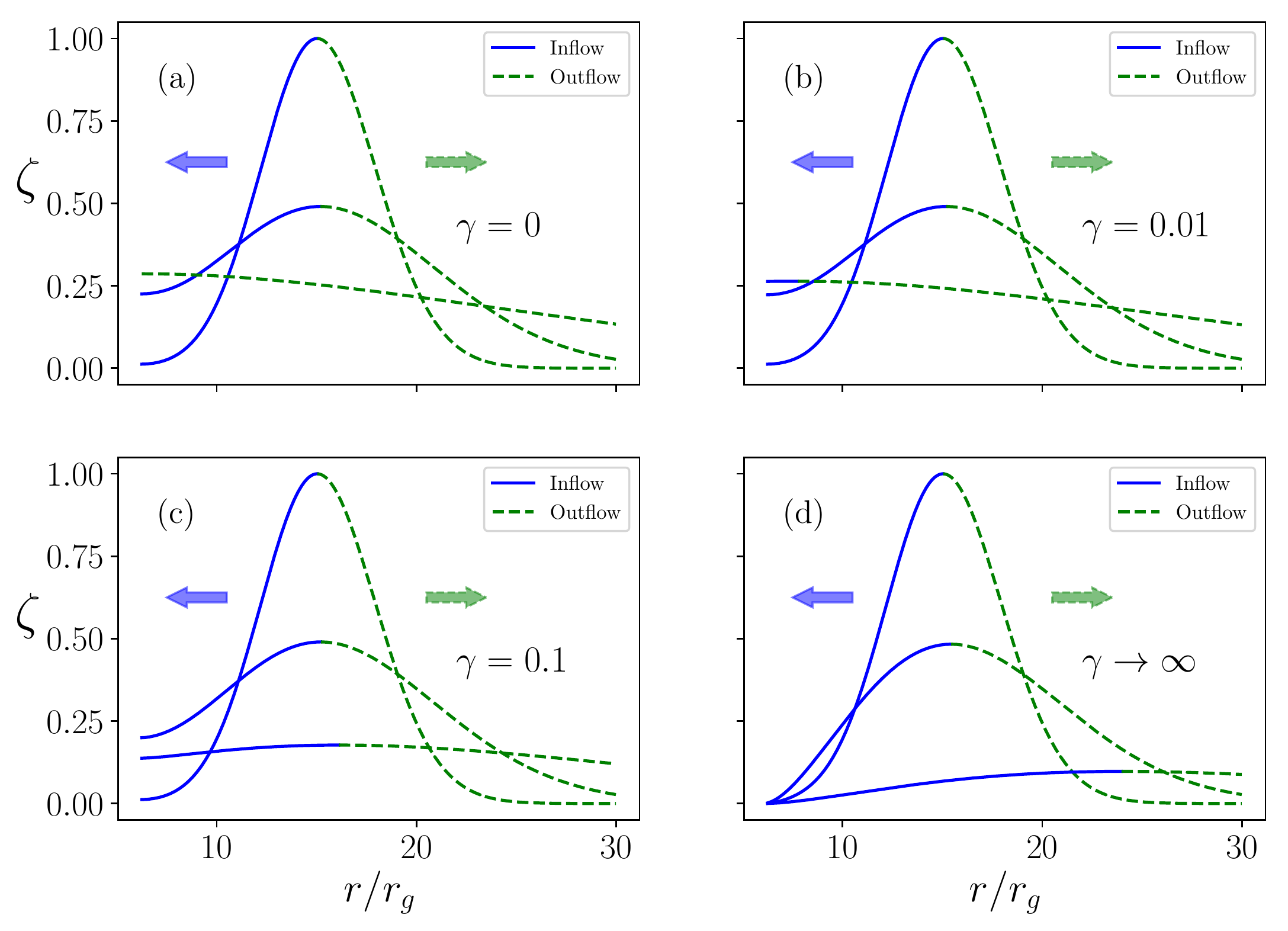}
\caption{Plots of the normalised dynamical variable $\zeta$ at three different times in the disc evolution, with varying disc parameters $\gamma$, labelled on each plot. Each plot contains curves produced at three different dimensionless times, $\tau = 0.17, 0.67, 5$, the later times can be identified by the decreasing peak magnitude of $\zeta$. The blue solid curves and green dashed curves represent regions in which the disc is inflowing and outflowing respectively. }
\label{zetacompare}
\end{figure*} 

\begin{figure}
\centering 
\includegraphics[width=.49\textwidth]{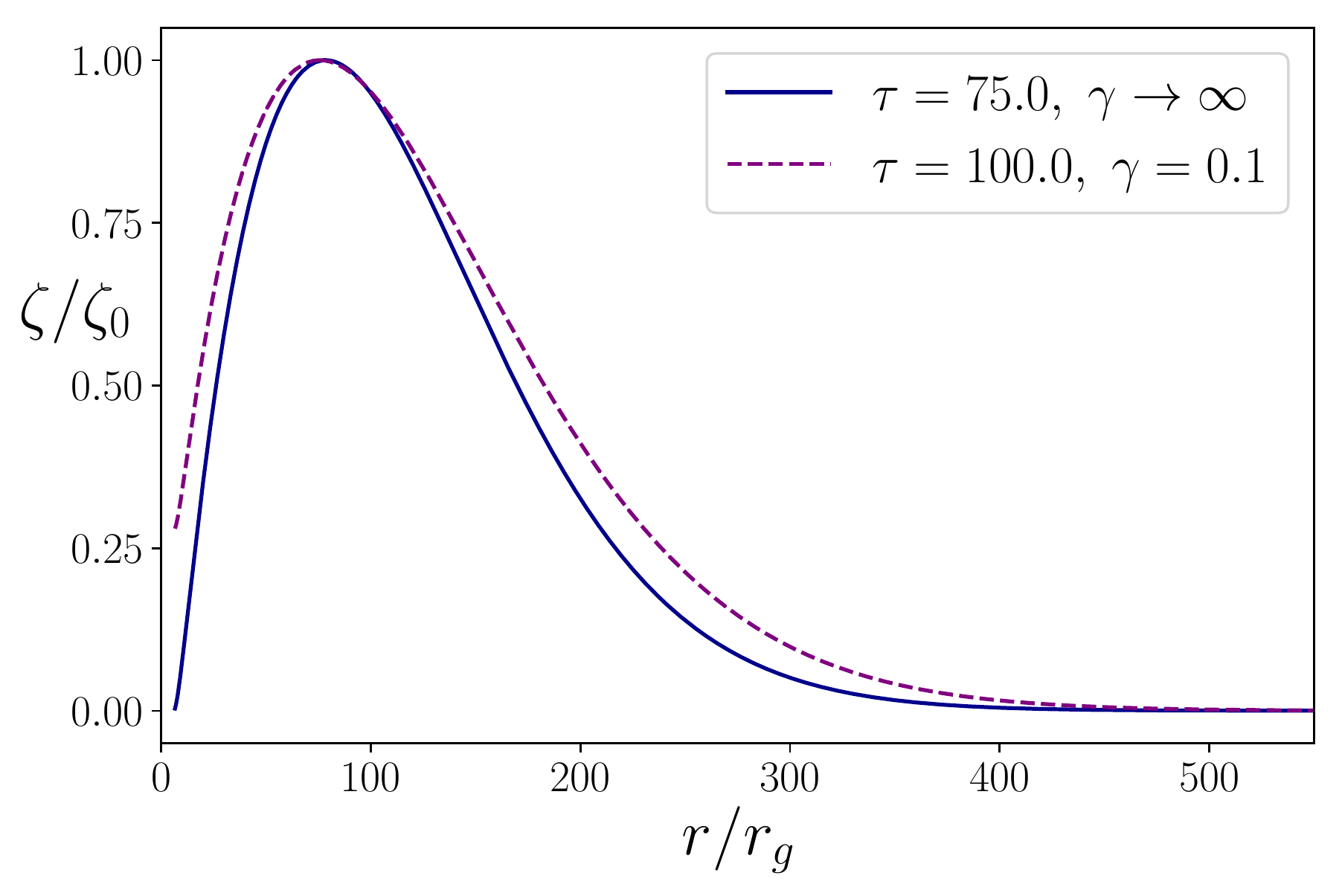}
\caption{Comparisons between a $\gamma = 0.1$ disc and a vanishing ISCO stress disc at very large times, both normalised by their {respective} magnitudes. Note both the different times at which the two curves are plotted (see text), and also the much enlarged radial range compared to Fig.\  (\ref{zetacompare}). }
\label{latetimecompare}
\end{figure}

 In paper I, the finding that $\zeta$ evolved to a global maximum at the ISCO for our simplest ($\gamma = 0$) finite stress disc models implied large scale radial outflow, a behaviour now seen as an extended transient state that a slightly modified angular momentum profile can eventually resolve.     Fig.\  (\ref{zetacompare}) shows explicitly the early to intermediate stages of the evolution of a set of discs, differing only in their respective  $\gamma$ parameters.   
Plots (a-d) in fig.\  (\ref{zetacompare}) each show three dimensionless times: $\tau = 0.17$, $ 0.67$ \& $5.0$.   
This corresponds to $56$, $221$ \& $1650$ days (respectively) in our fiducial model.    

The very different behaviour of the finite and vanishing stress discs may be seen by comparison of plots (5a) \& (5d), particularly at $\tau = 5$.  This very different $\zeta$ evolution is responsible for the shallow versus steep light curve behaviour described in BM18 and paper I.  As has been noted in paper I, the $\tau = 5$  curve in plot (a) exhibits permanent stalling, with outflow at all points.

The three finite stress curves (5a), (5b) \& (5c) are indistinguishable at the earlier times $\tau = 0.17$ and $0.67$, whilst they all differ from the vanishing stress plot (5d).  The reason for this is that at early times all three finite stress solutions are in the stalled phase of accretion, a phase which is vanishingly brief for the vanishing ISCO stress disc.  The stalled phase for the $\gamma = 0.01$ disc is extremely extended.   Over the entire time examined, the (a) \& (b) curves remain near indistinguishable, differing only in the innermost  $\sim 2 r_g$ exterior to the ISCO radius. 

By $\tau = 5$, however, the $\gamma = 0.1$ disc in plot (c) shows an intermediate type behaviour, qualitatively different from the extremes seen in plot (a) and plot (d).  The intermediate solution is transitioning from a completely stalled to a steady phase of accretion.  This is also seen in fig.\  (\ref{latetimecompare}), which shows a comparison between the very-late time behaviour of a $\gamma = 0.1$ disc and a vanishing ISCO stress disc.  By this time, the  $\gamma = 0.1$ disc solution is converging towards the profile of the vanishing ISCO stress solution, just as predicted by the analytical arguments of section \S\ref{newmodel}.   Note that the $\gamma = 0.1$ disc is plotted at a later time than the vanishing-ISCO-stress disc.   The need for this time offset is in fact due to the presence of an extended, intermediate time, stalled phase.

\subsection{Summary of numerical results} 

We have integrated the relativistic thin disc evolution equation (\ref{eq}), using a modified angular momentum profile of the form given by equation (\ref{angprofile}).  Solutions were found for both a vanishing ISCO stress and a finite ISCO stress, where the inner boundary condition in the latter case was determined by equation (\ref{gammaBC}), with a particular range of values of the $\gamma$ parameter motivated by 3D GRMHD simulations of thin discs. 
We have shown that the inclusion of non-ideal deviations from circular orbits near the ISCO breaks the late time luminosity fall-off dichotomy of the ``zeroth order" vanishing stress and finite ISCO stress solutions presented in paper I.  The $\gamma$ parameter in essence joins them in a continuous model parametrisation.   These new models demonstrate a richer behaviour with features of both the more idealised vanishing and finite stress ISCO stress solutions. 

The key finding of these numerical studies is that, rather than acting to permanently stall accretion, a finite stress at the ISCO causes the evolution of a relativistic thin disc to follow three distinct stages.   In common with vanishing stress discs, accretion with a finite ISCO stress begins with an initial phase of disc material flowing from the disc towards the ISCO, accompanied by an increase in the disc luminosity. This inrush of material leads to a build up of density in the inner regions.   In a disc with non-zero ISCO stress this leads to an extended, but finite,  period of stalled accretion.  During this interval,  the disc displays behaviour characteristic of the set of  ideal finite stress solutions derived in paper I -- most notably, significantly shallower light-curves than vanishing stress discs.  The gradual depletion of the inner disc from continued accretion then initiates the third stage of the accretion process, the approach to the steady state.  In this final stage of accretion, all discs, independent of their non-zero parameter $\gamma$, evolve in a manner nearly identical to that of a traditional vanishing stress solution.

Knowledge of $\gamma$ is needed to determine the duration of the stalled accretion phase. Dedicated  GRMHD simulations will be valuable in achieving this goal.

\section{simulation and observational support for stalled accretion}
\subsection{Comparison with numerical simulations}
We have seen that $\gamma$-disc evolution is comprised of three distinct stages.   By contrast, the idealised zero ISCO stress disc evolution is a two stage process.    The hallmark of a finite ISCO stress is the existence of an extended intermediate period of stalled accretion.   It is natural to ask whether this clear prediction of our one-dimensional model has actually been seen in 3D GRMHD simulations of accretion discs, where the stress is self-determined. 

The period of stalled accretion begins only after many viscous times, when the disc surface density has peaked in the inner near-ISCO regions.   Running full 3D GRMHD simulations out to such long times is presently prohibitively expensive,  and it seems unlikely that in the near future this will change dramatically.     However, current 3D simulations do at least provide an arena to test the {\em early time} behaviour of our model -- the initial spreading phase -- where the behaviour of our reduced models should be compatible with what is seen in 3D simulations.

\begin{figure}
\includegraphics[width=.5\textwidth]{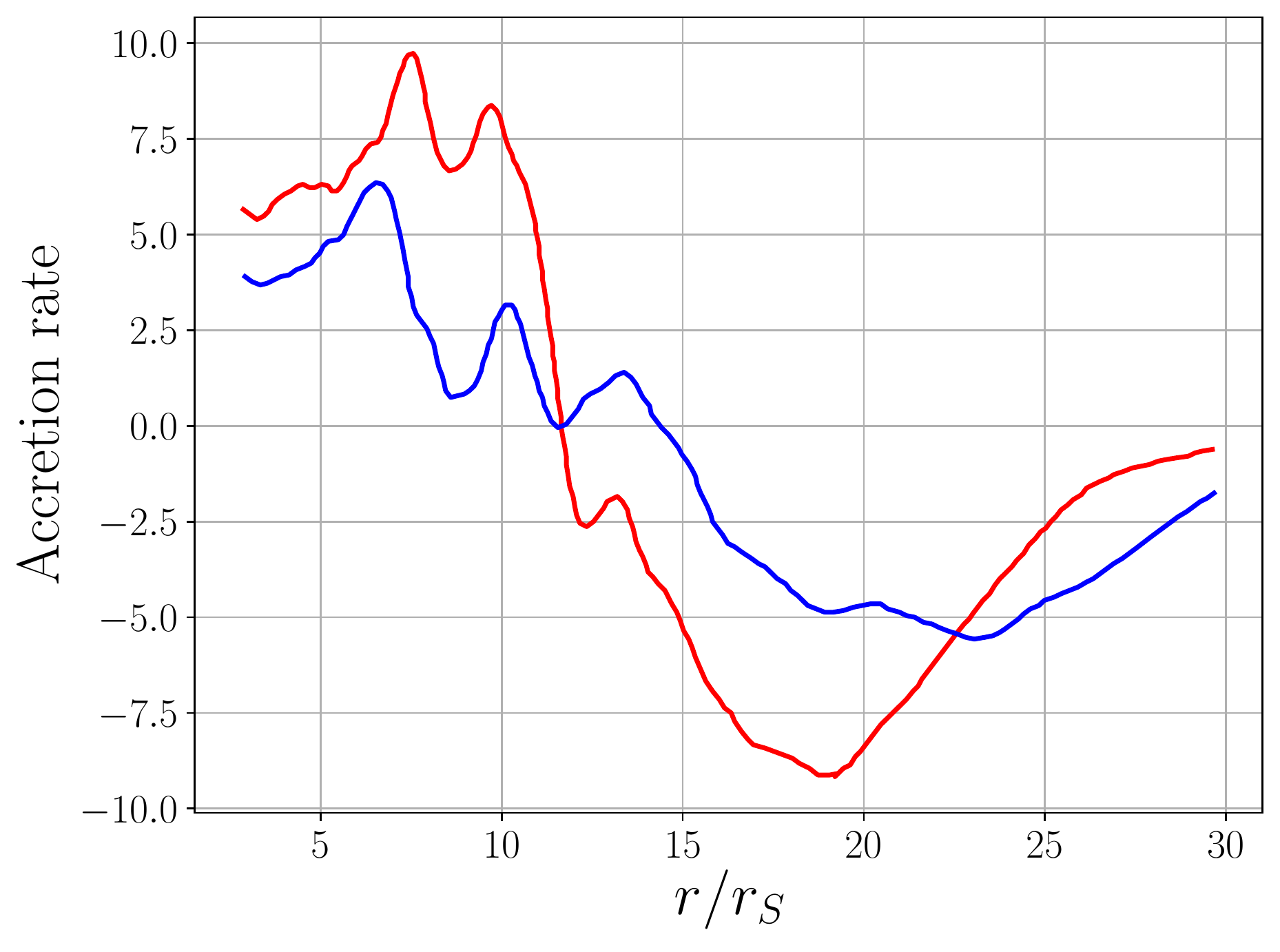}
\caption{Accretion rate versus radius taken from the 3D  MHD simulation of Hawley \& Krolik (2001), which uses a Paczy\'{n}ski-Wiita potential, at two different times. The early time profile is coloured red and correspnds to $t = 1000 \,GM/c^3$.   The blue curve corresponds to the later time $t = 1500\,GM/c^3$. }
 \label{earlyH}
\end{figure}

\begin{figure}
\includegraphics[width=.5\textwidth]{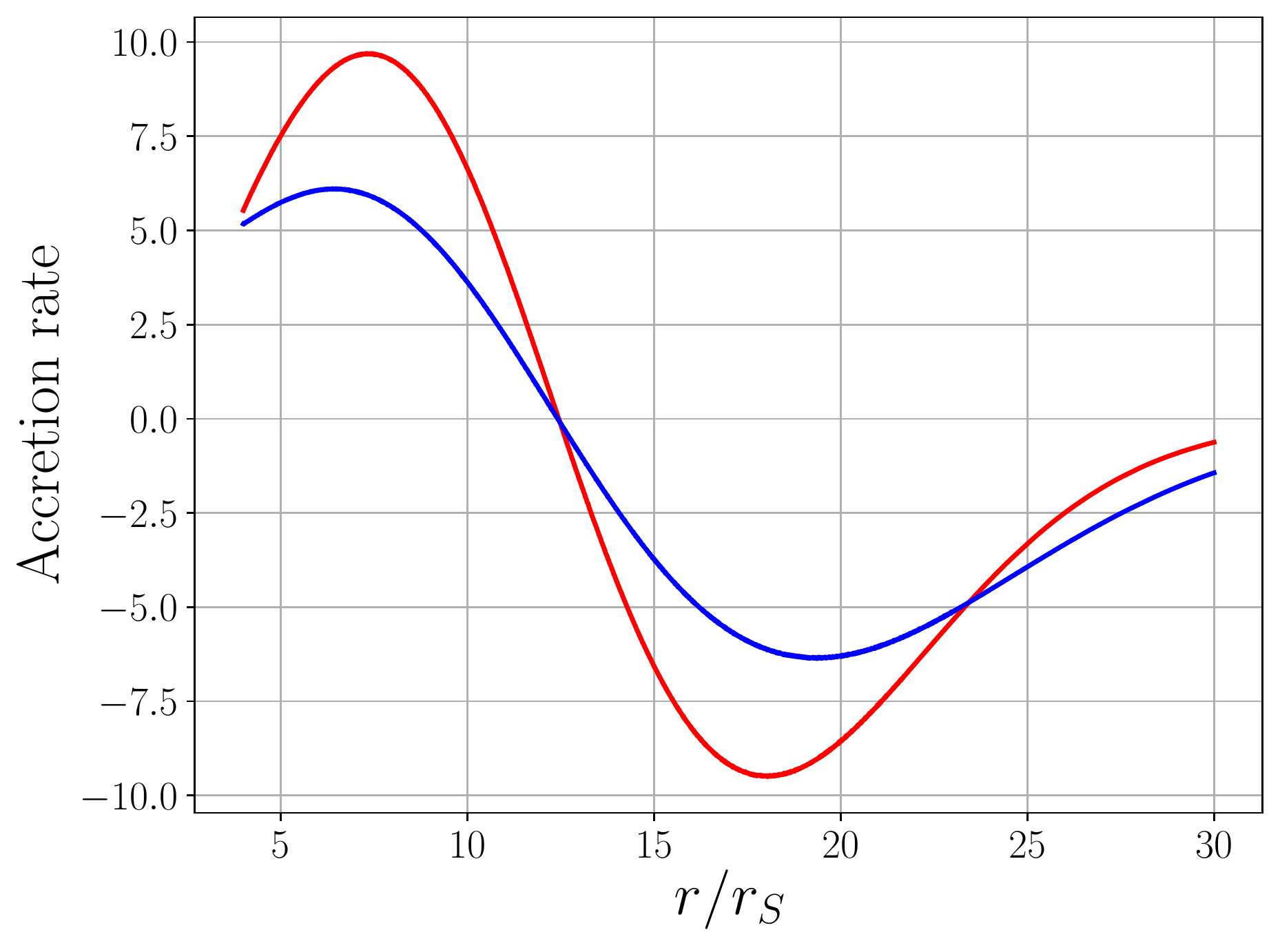}
\caption{Accretion rate versus radius for a 1 dimensional Schwarzschild disc model with a constant turbulent stress ($\W = w = $ constant; hence a finite stress at the ISCO) and $\gamma = 0.1$.  The magnitude of the accretion rate is normalised so as to have the same peak value at $t = 1000\, GM/c^3$ as the Hawley \& Krolik (2001) figure. The red curve corresponds to the earlier time, while the blue curve is the later time.   }
 \label{earlyM}
\end{figure}

A detailed quantitive comparison between our one dimensional disc models and the early time behaviour of 3D simulations will be the subject of a future study by the authors, tailored specifically to this task.    For the present,  we begin with a comparison of the evolution of the accretion rate as a function of radius in our 1D disc with a set of published 3D MHD simulations from Hawley \& Krolik (2001).    Fig.\  (\ref{earlyH}), taken from the study shows two radial accretion rate profiles and two different times.    By comparison, fig.\ (\ref{earlyM}),  shows the corresponding colour-coded profiles taken from a 1D disc model.  (Details are provided in the caption.)   The accretion rate profiles from our 1D disc model are encouragingly similar to those of the full 3D simulations, at least at these early times.  

Unlike 3D simulations, 2D axisymmetric simulations may be carried out to the much longer times required to enter the stalled phase of accretion.   The catch is that the calculations require the input, ``by hand'', of a sub-grid mean field dynamo.   (Recall that there is no true dynamo in 2D flow [Moffatt 1978].)    There is no obvious reason why this dynamically {\it ad hoc} procedure should alter our prediction of three stage accretion, however.     A series of twelve 2D GRRMHD simulations were performed by S\c{a}dowski \textit{et al}. (2015), five with initial near-Eddington accretion and seven with initial super-Eddington accretion.  The seven super-Eddington simulations are thick disc simulations ($H/R \sim 1$), and are therefore not appropriate for comparison with thin disc models.   However, the five starting with near-Eddington accretion rates have typical values of $H/R$ of $ \sim 0.1$, with a maximum value $H/R = 0.2$, and are therefore more amenable to comparison with our thin disc model predictions.  

The findings by S\c{a}dowski \textit{et al} are very revealing.   All five of their simulations followed an initial evolutionary pathway in which disc temperature and thickness evolved from their initial conditions towards values consistent with a turbulent accretion flow.   However, in each of the simulations, the accretion flow transitioned to a state in which the disc density peaked in a (vertically) narrow region near the ISCO.   This build-up was then followed by an almost immediate pronounced drop in the accretion rate, despite the the vigorous presence of a fully resolved MRI.  In other words, the accretion in these discs was {\em stalled} by a build-up of disc density in the inner disc regions, much as predicted by our finite ISCO stress one dimensional disc models.   A dedicated study of ``aided'' 2D GRMHD disc simulations would be of great interest here to establish more definitively the existence of disc stalling and whether, at yet longer times, accretion is restored.  

\subsection{TDE light-curves}
The possibility of observing luminous X-ray flares resulting from the disruption of a wayward star by the gravitational tides of a supermassive black hole was first described over 30 years ago (Rees 1988, Hills 1975).   It is only quite recently,  however, that extended observations of tidal disruption events have been made for several years following the initial flare (Van Velzen \textit{et al.} 2019, Auchettl, Guillochon \& Ramirez-Ruiz 2017).  The late time evolution of the luminosity emergent from such events is of particular interest, because the dominant component may well be produced by an accreting debris disc. late time TDE observations therefore act as an interesting observational diagnostic for testing accretion disc models.   One particularly useful model diagnostic is the late time luminosity decay index $n$, in $L \sim t^{n}$. 

There are at least three competing late time TDE models.   First, there is the original Rees (1998) `fallback' model, which assumes equal mass in equal energy intervals (and ballistic dynamics), leading to $n = - 5/3$. Second, there are disc models.  The predictions of disc accretion vary, depending on how exactly the turbulent stress ($\W$) is parameterised.  Most typical is a constant $\alpha$ (Shakura \& Sunyaev 1973) model with a disc opacity dominated by electron scattering.  In this case,  a vanishing stress disc model (Cannizzo \textit{et al.} 1990) extends the duration of the emission somewhat, with a typical index of $n = - 1.19$.   Finally, the  $\gamma$-disc model investigated here, again assuming electron scattering opacity, has an even shallower luminosity fall-off, with an index around $n = -0.79$ (paper I) in the stalled accretion regime, appropriate for comparison to TDE observations (Fig.\  \ref{n_compare}).   

A recent compilation of the late time power law indices of X-ray TDE sources (Auchettl, Guillochon \& Ramirez-Ruiz 2017) is summarised in Table \ref{table2}.   At present, only four ``confirmed'' X-ray TDEs are currently available for comparison, and the light curves of the sources are sparsely sampled.  (The power law indices should be relied on to no more than one significant figure.)    With these caveats, it is striking that {\em none} of the TDE power law indices are greater than 1 in magnitude.    This is exactly what would be expected for a simple finite ISCO stress disc model, and is a firm prediction. 

It is important to exert caution when fitting model predictions of the behaviour of the \textit{bolometric} luminosity to the observed luminosity within any finite-frequency-width band. We will demonstrate in a forthcoming paper (Mummery \& Balbus 2019, in preparation) that the observed temporal behaviour of transient accretion discs can be strongly observational-band dependent, and present analytical expressions for the time dependent luminosity in Optical, Ultra-violet and X-ray bands.  While these results are of course somewhat more complicated than the simple bolometric luminosity power laws, all bandpasses involve the index $n$ used here, so competing disc models may be tested and distinguished. 

Van Velzen \textit{et al}. (2019) present late time observations of TDEs at far ultra-violet (FUV) wavelengths.  Their FUV light-curves are in fact consistent with that produced by accretion from a thin disc,  and appear to be inconsistent with the Rees (1988) model.  At FUV wavelengths,  it is difficult to differentiate between vanishing and finite ISCO stress disc models, in fact Van Velzen \textit{et al}. (2019) use a purely Newtonian disc model. The major difference in the gross disc structures is in the inner hot disc regions, therefore observations at X-ray wavelengths provide the best data with which to differentiate between the two disc models.  To calculate the observed X-ray flux emerging from the innermost regions of accretion discs is best approached with a relativistic disc model, as we have described in this work. 

We suggest that these early inferences from TDE observations are at the very least suggestive of stalled accretion, and thus a non-zero ISCO stress.  The possibility of using future TDE observations to extract further details of the inner disc stress structure is a viable and  exciting proposition: the nature of the stress at the ISCO has long been a controversial question.   TDEs offer what may prove to be a powerful observational discriminator.    If future observations of TDEs consistently find late time power law indices smaller than unity in magnitude (with bandpass restrictions taken into account), this would provide strong evidence in favour of a finite ISCO stress, as well as elucidate the physical origins of these remarkable sources.

\begin{table}
\centering
\begin{tabular}{|c|c|}
\hline
Rees (1988) & $-1.67$ \\ \hline
Cannizzo \textit{et al.} (1990) & $ -1.19$ \\ \hline
Mummery-Balbus (2019) & $-0.79$ \\ \hline
\end{tabular}
\caption{Model predictions of late time TDE bolometric luminosity decay indices}
\label{table1}
\end{table}

\begin{table}
\centering
\begin{tabular}{| l | c |}
\hline
ASASSN-14li & $-1.0$ \\ \hline
 Swift J1644+57 & $-0.71$ \\ \hline
Swift J2058+05& $-0.16$ \\ \hline
 XMMSL1 J0740-85 & $-0.75$ \\ \hline 
\end{tabular}
\caption{ The four well-observed sources from AGR (left) and their deduced late time luminosity power law index (right).}
\label{table2}
\end{table}

\section*{Acknowledgments} 
This work is partially supported by STFC grant ST/S000488/1.  It is a pleasure to acknowledge useful conversations with  M.\ Begelman,  J.\ Krolik, B.\ Metzger, and W.\ Potter.


\label{lastpage}
\end{document}